\documentclass[letterpaper,showpacs,preprintnumbers,amsmath,amssymb,nofootinbib,twocolumn,superscriptaddress,notitlepage,prl]{revtex4-1}
\usepackage{graphicx}
\usepackage[usenames,dvipsnames]{color}
\usepackage[colorlinks,citecolor=blue]{hyperref}
\usepackage{amsmath}
\usepackage{bbm}
\usepackage{amsfonts}
\usepackage{amssymb}
\usepackage{latexsym}
\usepackage{graphicx}
\usepackage[english]{babel}
\usepackage{multirow}
\usepackage{float}
\usepackage{url}
\usepackage{slashed}
\usepackage{xcolor} 

\newcommand{\be}{\begin{equation}}
\newcommand{\ee}{\end{equation}}
\newcommand{\ba}{\begin{array}}
\newcommand{\ea}{\end{array}}
\newcommand{\bea}{\begin{eqnarray}}
\newcommand{\eea}{\end{eqnarray}}

\newcommand{\eecca}{$e^+ e^- \to \chi\bar \chi \gamma$\,}

\newcommand{\GeV}{\,{\rm GeV}}

\usepackage{ulem,fancyvrb}
\usepackage{xcolor}

\begin{document}
\title{Probing millicharge at BESIII}

\author{Zuowei Liu}\email{zuoweiliu@nju.edu.cn}
\affiliation{Department of Physics, Nanjing University, Nanjing 210093, China}
\affiliation{Center for High Energy Physics, Peking University, Beijing 100871, China}
\affiliation{CAS Center for Excellence in Particle Physics, Beijing 100049, China}

\author{Yu Zhang} \email{dayu@nju.edu.cn}
\affiliation{Department of Physics, Nanjing University, Nanjing 210093, China}
\affiliation{CAS Center for Excellence in Particle Physics, Beijing 100049, China}

\begin{abstract}

We propose to search for millicharged particles at the BESIII detector 
which is operated at the Beijing Electron Positron Collider. 
We compute the monophoton signal events at the BESIII detector due 
to millicharged particle production, as well as due to standard model 
irreducible/reducible backgrounds. 
By utilizing all the data accumulated at the BESIII detector since 2011, 
we derive new leading upper limits on millicharge,  
$\varepsilon \lesssim (0.86-2.5) \times10^{-3}$, for the mass range, 
$0.1~\text{GeV} \lesssim m  \lesssim 1~\text{GeV}$. 
Furthermore, projections with more data to be collected at the BESIII detector 
are also made. 
Our analysis significantly reduces the parameter region of millicharge 
to account for the anomalous 21 cm absorption signal near redshift $z\simeq 17$ 
{recently} observed by the EDGES experiment.

\end{abstract}

\maketitle

{\it Introduction.}
Electric charge quantization is an empirical fact. 
Charge quantization is also theoretically related to the magnetic monopole, because 
if any magnetic monopole exists in the Universe, it  
quantizes the electric charge \cite{Dirac:1931kp}. 
However, so far, 
there is no clear experimental evidence to support 
the existence of the magnetic monopole. 
Thus we do not know yet what mechanism 
leads to electric charge quantization.


{A number of experiments 
have been carried out to detect the 
non-integer charge of particles in the standard model (SM), 
and very stringent limits have been obtained.} 
For example, 
the charge of hydrogen atom and neutron are 
measured to be smaller than $10^{-21}e$, 
where $e$ is the magnitude of the electron charge 
\cite{Marinelli:1983nd, Bressi:2011yfa, 
Baumann:1988ue, Siemensen:2018cjm}. 
There are also searches for 
new particles beyond the standard model that 
carry electric charge, and very strong constraints 
on electrically charged new particles have been imposed from various 
laboratory experiments and astrophysical processes 
(see e.g.\ 
\cite{Davidson:1993sj, Davidson:2000hf, Jaeckel:2010ni} 
for the review on the constraints). 
The electrically charged particles beyond SM are referred as 
millicharged (or minicharged) particles, 
{since usually only new particles with very small electric charge are allowed.}

To parameterize the extremely weak coupling between a millicharged 
fermion and the SM photon, we employ the following 
interaction Lagrangian 
\be
{\cal L}_\text{int} = e \varepsilon A_\mu \bar \chi \gamma^\mu \chi, 
\label{eq:millicharge}
\ee
where $\chi$ is the millicharged particle, 
$A_\mu$ is the SM photon, 
and $\varepsilon$ is the millicharge (normalized to 
the magnitude of the electron charge). 
There are viable theoretical models in which millicharged particle can naturally occur. 
For example, millicharge particles may be present in models in which a kinetic mixing 
term is introduced between different $U(1)$ gauge fields 
\cite{Holdom:1985ag,Holdom:1986eq}. 
Millicharged particles can also arise in Stueckelberg extensions of the 
standard model in which mass terms generated by the Stueckelberg mechanism 
mix the SM $U(1)_Y$ gauge boson and new Abelian gauge bosons 
in the hidden sector beyond SM 
\cite{Cheung:2007ut, Feldman:2007wj}. 
In this paper, we only consider the millicharged particle 
and the interaction given by Eq.\ (\ref{eq:millicharge}); 
we decouple all other particles that appear in a specific 
model and do not consider any additional interaction 
between millicharged particles with SM.


Recently, an anomalous absorption signal near redshift $z \simeq 17$ 
in the cosmological 21 cm spectrum was observed by the 
EDGES experiment \cite{Bowman:2018yin}. 
To explain such a signal, a number of 
papers have used millicharged dark matter (DM) particles   
to cool the hydrogen atom in the Universe. 
Millicharged DM is a good candidate to explain the 21 cm anomaly, 
since the interaction cross section between DM and baryons 
exhibits a $v^{-4}$ behavior which is consistent with cosmological 
observations. 
Our analysis in this paper has direct implications to the 
parameter space of millicharged particles that can explain the 
21 cm anomaly 
\cite{Munoz:2018pzp, Berlin:2018sjs, Barkana:2018qrx}.


In this paper, we propose to search for millicharged particles below GeV 
at the BESIII detector 
which is operated at the Beijing Electron Positron Collider (BEPCII). 
The existing laboratory constraints for MeV-GeV millicharged particles include 
bounds from the SLAC electron beam-dump experiment \cite{Davidson:1991si}, 
bounds from the SLAC MilliQ searches \cite{Prinz:1998ua},  
bounds from the E613 experiment \cite{Soper:2014ska}, 
and bounds from MiniBooNE \cite{Magill:2018tbb}. 
Recently, CMS collaboration \cite{CMS:2012xi} excludes 
particles with electric charge $2e/3$ ($e/3$) below 310 (140) GeV. 
There is also a proposed experiment at the LHC aiming to 
detect millicharged particles \cite{Haas:2014dda}. 
{Here we use the monophoton signal to probe the MeV-GeV millicharged particles
by analyzing the BESIII data of $\sim$15/fb.}
The monophoton signature has been considered previously 
in DM searches at $e^+ e^-$ colliders 
\cite{Borodatchenkova:2005ct, Zhu:2007zt, Fayet:2007ua, 
Essig:2009nc, Yu:2013aca, Essig:2013vha}. 
Here we first carry out a detailed analysis at the BESIII detector 
by taking into account various backgrounds.
We derive the BESIII sensitivity to the millicharge, 
and show that the BESIII detector can probe the parameter region 
that has not been constrained by previous experiments.


{\it Signals of millicharged particles at electron colliders.} 
Millicharged particle can be produced at particle colliders 
via its coupling with the standard model photon.  
However, if the millicharge is very small, 
the produced millicharged 
{particle is often undetectable in practice, 
because only a feeble signature inside  
particle detectors could be produced.
Thus one relies on final state particles 
produced in additional to the millicharged particle 
for the detection, which is analogous 
to most dark matter searches at particle colliders.}

\begin{figure}[htbp]
\vspace{0.2cm}
\centering
\includegraphics[scale=1]{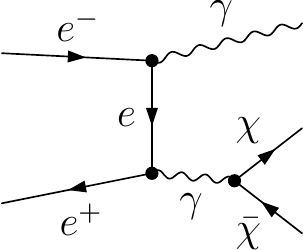}
\caption{Feynman diagram for the process \eecca. 
The diagram with photon radiated by the positron is 
included in the analysis, but not drawn here. 
The diagrams with photon radiated by millicharged 
particles are not considered since 
they are suppressed by $\varepsilon$.}
\label{fig:eecca}
\end{figure}

At the electron-positron collider, we use the monophoton 
signal to search for the millicharged particles. 
The Feynman diagram for the production process of the single photon 
in association with millicharged particles, \eecca, 
 is shown 
in Fig.\ (\ref{fig:eecca}), where $\chi$ stands for the millicharged particle. 

The differential cross section for 
the $e^+ e^- \to \chi \bar \chi \gamma$ process is given by 
\be
{d \sigma \over d E_\gamma d z_\gamma} =
 {8 \alpha^3 \varepsilon^2 (1 + 2 m_\chi^2/s_\gamma)  \beta_\chi 
 \over 3 s E_\gamma (1 - z_\gamma^2)}  
 \Bigg[1  + {E_\gamma^2 \over s_\gamma}(1+z_\gamma^2) \Bigg], 
 \label{eq:np}
\ee
where $E_\gamma$ is the energy of the final state photon, 
$z_\gamma \equiv \cos\theta_\gamma$ with 
$\theta_\gamma$ being the relative angle between 
the final state photon and 
the beam direction of the initial state electron,  
$s$ is the square of the center-of-mass energy,  
$m_\chi$ is the mass of the millicharged particle, 
$s_\gamma = s - 2\sqrt{s} E_\gamma$, 
and $\beta_\chi = (1-4m_\chi^2/s_\gamma)^{1/2}$. 
Here we have integrated over all possible 
momenta for the two final state millicharged particles 
and neglected the electron mass.

\begin{figure}
\begin{center}
\includegraphics[angle=0,width=3.2in,height=2.4in]{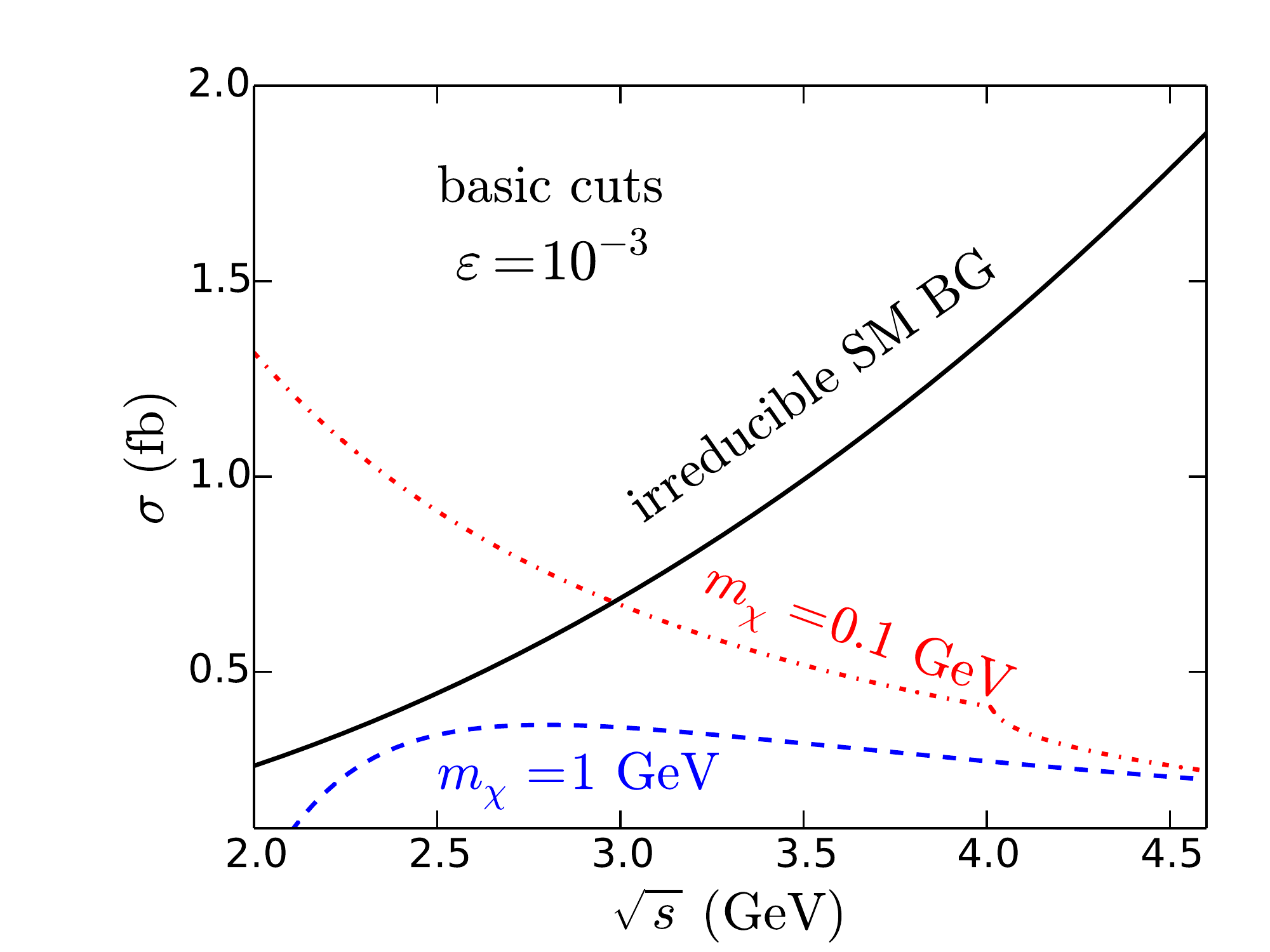}
\caption{The monophoton cross sections of the irreducible SM background and 
of millicharged particles as a function of $\sqrt s$ in 
the BEPCII energy range, $2.0 \,{\rm GeV} \le \sqrt s \le 4.6 \,{\rm GeV}$. 
The kink in the $m_\chi = 0.1$ GeV curve is due to the 
maximum energy measured in the EMC at BESIII which is 2 GeV. 
}
\label{fig:xsec}
\end{center}
\end{figure}

{\it The irreducible background.} 
The major irreducible standard model background processes to the monophoton 
signal at the electron-positron collider are the 
$e^+ e^- \to \nu_\ell \bar \nu_\ell \gamma$ processes, 
where $\nu_\ell = \nu_e, \nu_\mu, \nu_\tau$ are the three standard model 
neutrinos. 
\begin{figure}[htbp]
\vspace{0.2cm}
\centering
\includegraphics[scale=0.9]{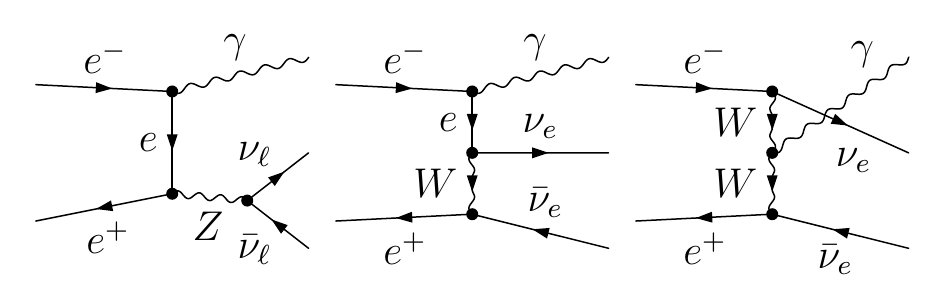}
\caption{Feynman diagrams for the process 
$e^+ e^- \to \nu_\ell \bar \nu_\ell \gamma$, 
where $\nu_\ell=\nu_e,\nu_\mu,\nu_\tau$ 
\cite{Gaemers:1978fe,Araki:2017wyg}.
}
\label{fig:eevva}
\end{figure}
The corresponding Feynman diagrams are displayed in Fig.\ (\ref{fig:eevva}). 
For electron neutrinos, both the $Z$-boson and the $W$-boson diagrams contribute; 
whereas the muon and tau neutrinos are only produced via the $Z$-boson 
diagrams. For electron colliders operated at the GeV scale, the 
diagram mediated by two $W$ bosons are significantly smaller than 
those with a single $W$ or $Z$ mediator. Thus we do not consider the 
diagram with two $W$ mediators for our analysis in the electron 
colliders running with GeV beam energy. 
The differential production cross section 
for the $e^+ e^- \to \nu \bar\nu \gamma$ processes 
mediated by a single $W/Z$ boson is given by 
\cite{Ma:1978zm} \cite{Gaemers:1978fe}
\be
{d\sigma \over d E_\gamma d z_\gamma}
=  {\alpha G_F^2  s_\gamma^2  
\over 4 \pi^2  s E_\gamma (1 - z_\gamma^2) } f(s_W)
\Bigg[ 1 + {E_\gamma^2 \over s_\gamma} (1 + z^2_\gamma)
\Bigg], 
\label{eq:irbg}
\ee
where $G_F$ is the Fermi constant, 
$s_W \equiv \sin\theta_W$ with 
$\theta_W$ being the weak  mixing angle, and 
$f(s_W) = 8 s_W^4- 4 s_W^2/3+1$. 
Here we have integrated over the momenta of the 
final state neutrinos and summed all three 
neutrino flavors. 
As shown in Fig.\ (\ref{fig:xsec}), 
the monophoton cross section due to the irreducible SM background grows with 
the colliding energy; 
however, the monophoton cross section due to millicharged particle production
 in the $m_\chi = 0.1$ GeV case increases when the colliding energy 
 {decreases.}
 Thus, electron collider with smaller colliding energy has a better sensitivity 
 {to kinematically accessible millicharged particles.}

There are other irreducible SM backgrounds due to 
semi-invisible meson decays; for example the decay mode 
$J/\psi \to \nu \bar \nu \gamma$ contributes to the irreducible background 
if the colliding energy is tuned to coincide with the mass of the $J/\psi$ meson. 
However, the branching ratio of these decay modes are typically very 
small, for instance, BR$(J/\psi \to \nu \bar \nu \gamma) = 0.7\times 10^{-10}$  
in SM \cite{Gao:2014yga}. Thus we exclude those 
irreducible {backgrounds in} meson decays in our analysis.


{\it Reducible backgrounds at the BESIII detector.}
Next we want to investigate 
the {reducible backgrounds} at the BESIII detector, 
which is located at the double-ring 
BEPCII with the beam energy ranging 
from 1.0 GeV to 2.3 GeV \cite{Asner:2008nq}. 
{The reducible backgrounds are present due to the 
limited detection capability of the BESIII subdetectors.}
The main drift chamber (MDC), the innermost sub-detector of BESIII, 
that determines the momentum of a charged particle, 
covers the polar angle $|\cos\theta|<0.93$ \cite{Asner:2008nq}. 
The electromagnetic calorimeter (EMC) that measures the energies 
and positions of electrons and photons consists of 
the barrel with angle coverage $|\cos\theta|<0.83$ and 
the endcap with angle coverage $0.85<|\cos\theta|<0.93$ \cite{Asner:2008nq}. 
The Time-of-Flight (TOF) sub-detector which is placed between the 
drift chamber and the electromagnetic calorimeter measures 
the flight time of charged particles. The TOF consists of 
the barrel with angle coverage $|\cos\theta|<0.83$ and 
the endcap with angle coverage $0.85<|\cos\theta|<0.95$ \cite{Asner:2008nq}.

\begin{figure}[htbp]
\vspace{0.2cm}
\centering
\includegraphics[scale=0.9]{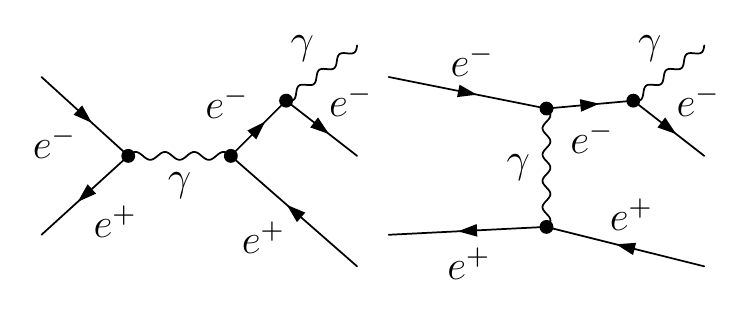}
\caption{Feynman diagram for the process $e^+ e^- \to e^+ e^- \gamma$. 
Here diagrams mediated by a virtual $Z$ boson are neglected.}
\label{fig:eeeea}
\end{figure}

{One of the most important reducible backgrounds}    
arises from the radiative Bhabha scattering, 
$e^+ e^- \to e^+ e^- \gamma$, where neither of the two final state electrons is 
detected. The Bhabha scattering can be mediated by either an s-channel or a t-channel 
virtual photon; because any of the four external fermion legs can radiate a photon, 
there are eight diagrams in the radiative Bhabha scattering. 
Two of the radiative Bhabha diagrams are shown in 
Fig.\ (\ref{fig:eeeea}). 

%

We further consider the $e^+ e^- \to \bar f f \gamma$ background where 
the final state $f$ fermion ($f \neq e$) {escapes} detection,  
and the $e^+ e^- \to \gamma \gamma  \gamma$ 
process where only one of the three final state photons is detected.
There are also reducible backgrounds from meson decays in which  
a final state photon is accompanied by several other 
particles going along beam directions.  
We consider all these reducible {backgrounds} in our analysis.


{\it Detector simulation.} 
We simulate signal events and different SM 
background events for various BESIII running energies, 
which are shown in Table (\ref{tab:data}). 
For the signal process $e^+ e^- \to \chi \bar \chi \gamma$ 
and the irreducible background $e^+ e^- \to \nu \bar \nu \gamma$, 
we generated one million points from the analytic differential 
cross sections, Eq.\ (\ref{eq:np}) and Eq.\ (\ref{eq:irbg}), 
in the $E_\gamma$-$z_\gamma$ plane using Monte-Carlo methods; 
each point has a weighted differential cross section. 

There are several singularities associated with the radiative Bhabha scattering. 
For the final state photon, neither infrared divergence nor the collinear singularity 
is relevant since the photon has to be detected.
However, because the final state $e^\pm$ has to be 
undetected in order to contribute to the background, the collinear singularity ($e^\pm$ 
going along the beam directions) cannot be removed.
This arises in the t-channel photon processes as shown in Fig.\ (\ref{fig:eeeea}). 
Although the electron mass makes this collinear singularity finite, 
it is still difficult to numerically compute the scattering cross section 
in the collinear region of the phase space which, however, dominates the total cross section 
\cite{Mana:1986je,Tobimatsu:1989yv,Tobimatsu:2001kb,Actis:2009uq}. 
We use {\sc FeynArts} \cite{Hahn:2000kx} and {\sc FormCalc} \cite{Hahn:1998yk} packages 
to numerically evaluate the cross section for the process $e^+ e^- \to e^+ e^- \gamma$ 
where the final state $e^\pm$ 
has $|\cos\theta|>0.95$ which is beyond the coverage of the TOF, 
as well as MDC and EMC.  
{\sc FeynArts} and {\sc FormCalc} packages 
output weighted scattering cross sections for different phase space points. 
We found that {\sc MadGraph} \cite{Alwall:2014hca} 
cannot sample the phase space efficiently due to the collinear singularity. 
We carry out similar calculations for 
$e^+ e^- \to \mu^+ \mu^- \gamma$ and 
$e^+ e^- \to \gamma\gamma \gamma$.  
We impose $E_\gamma>1$ MeV \cite{cut}
to remove the infrared divergence 
in $e^+ e^- \to \gamma \gamma \gamma$.

We use {\sc EvtGen} \cite{Lange:2001uf,Ping:2008zz} 
to simulate reducible background coming from meson decays.  
We generate $2\times10^8$ meson decay events for $J/\psi$, 
$10^8$ events for $\psi(3683)$, 
$10^8$ events for $\psi(3770)$, 
and $3\times10^7$ events for $\psi(4040)$.

The energy and position information of photon and electron are determined by 
the EMC. The energy resolution of the EMC 
{at the BESIII detector} 
is \cite{Asner:2008nq} 
\be
\sigma(E)/E=2.3\%/\sqrt{E/{\rm GeV}}\oplus 1\%. 
\label{eq:deve}
\ee
The angular resolution also depends on the energy of the particle; 
we provided the following fitted function  
which gives a nice approximation to the angular resolution \cite{Prasad:2015bra}
\be
\sigma(\theta)= (0.024/\sqrt{E/\GeV}-0.002) ({\rm rad}). 
\label{eq:devtheta}
\ee
To simulate the detector effects on the final state particles, 
we smear the energy and the polar angle for the 
final state electron and photon using Gaussian distributions 
which take into account the resolution functions, 
Eq.\ (\ref{eq:deve}) and Eq.\ (\ref{eq:devtheta}).


{\it Detector cuts.}
The energy measurement for electrons or photons in the EMC 
{at the BESIII detector} 
ranges from 20 MeV to 2 GeV. 
{We follow the cuts used by the BESIII Collaboration} \cite{Ablikim:2017ixv}: 
(hereafter the basic cuts)
photon candidates must 
satisfy 
$E > 25 $ MeV in the barrel ($|\cos \theta|<0.8$)
or $E > 50 $ MeV in the end-caps ($0.86<|\cos\theta|<0.92$).

\begin{figure}[htbp]
\vspace{0.2cm}
\centering
\includegraphics[scale=0.5]{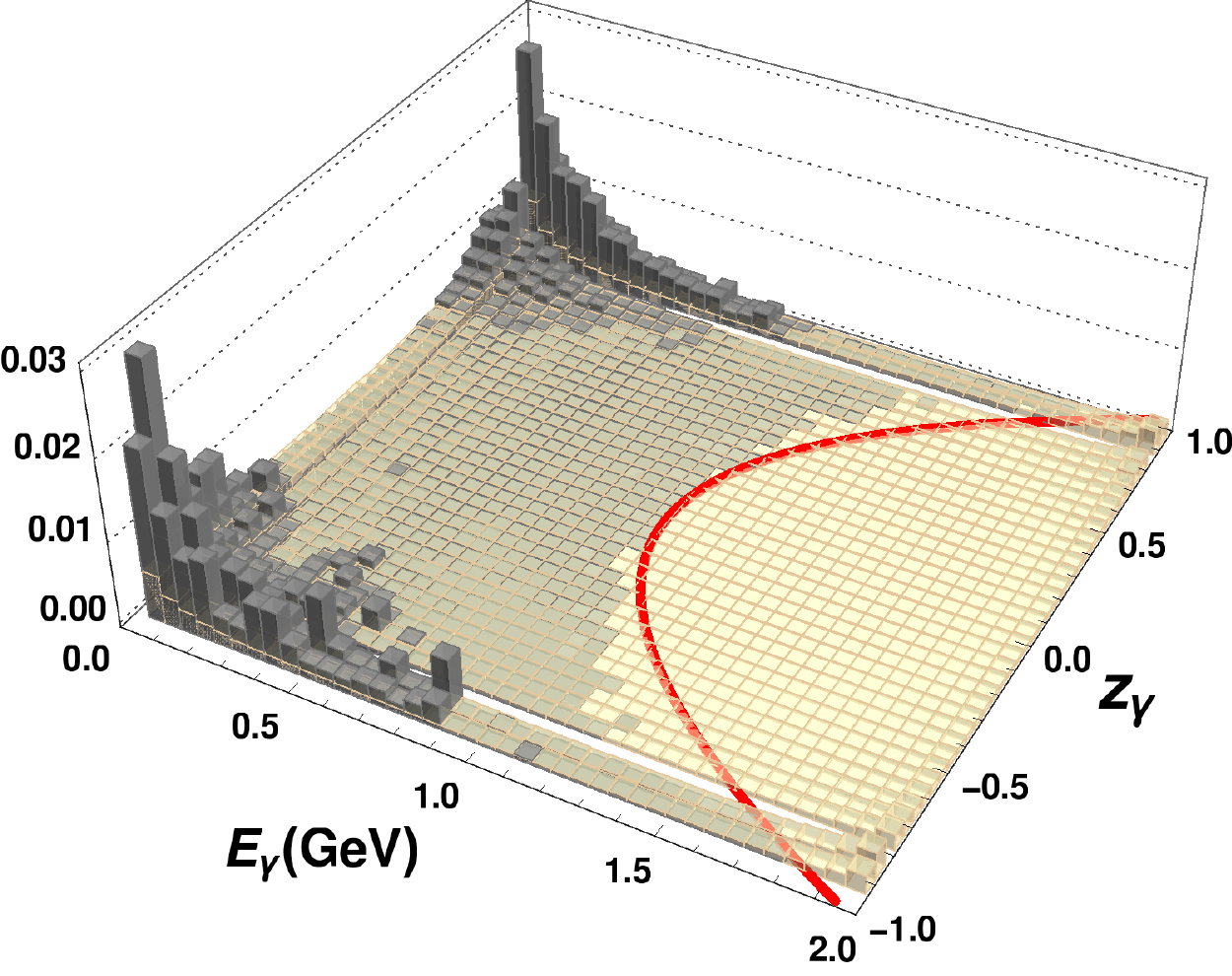}
\caption
{Photon $E_\gamma-z_\gamma$ normalized distributions 
in $e^+ e^- \to \chi \bar\chi \gamma$ 
with $m_\chi=0.1\, \GeV$ 
{(yellow)}
and in $e^+ e^- \to e^+ e^- \gamma$ (gray) 
at $\sqrt{s}=4.18$ GeV. 
The red curve is 
$E_\gamma/\text{GeV}=0.99 z_\gamma^2 + 0.99$.}
\label{fig:dis}
\end{figure}

However, after the basic cuts, the reducible background due to 
$e^+ e^- \to e^+ e^- \gamma$ is still very large, as shown in 
Fig.\ (\ref{fig:dis}) where we display 
$E_\gamma-z_\gamma$ normalized distributions for 
both the monophoton events 
due to millicharged particle production process and 
the $e^+ e^- \to e^+ e^- \gamma$ process. 
To suppress the big reducible backgrounds, we further impose the following 
cuts (hereafter the advanced cuts) on top of the basic cuts:
\be
E_\gamma/\text{GeV}> a\, z_\gamma^2 + b, 
\ee
where $a$ and $b$ are free parameters to be fixed by {maximizing} 
the significance. 
The advanced cuts are motivated by the fact that the final photon 
in the central region ($|\cos\theta|\ll 1$) cannot have a sufficient large energy due to 
energy conservation in the $e^+ e^- \to e^+ e^- \gamma$ process 
with both final $e^\pm$ going along the beam directions. 
The 
$e^+ e^- \to \mu^+ \mu^- \gamma$ and  
$e^+ e^- \to \gamma \gamma \gamma$ 
background processes exhibit similar distributions. 
As shown in Table (\ref{tab:cuts}) and in Fig.\ (\ref{fig:dis}), 
the advanced cuts are 
very efficient in eliminating these reducible {backgrounds}.

\begin{table}[h] 
\begin{center} 
\begin{tabular}{|c|c|c|c|c|c|c|} 
\hline 
\hline
Cuts & $\chi\bar{\chi}\gamma$  & $\nu\bar{\nu}\gamma$& ${e}^+{e}^-\gamma$ 
&$\mu^+\mu^-\gamma$&$\gamma\gamma\gamma$ &${\cal S}$ \\
\hline
Basic &32.3&1.39&$6.9\times10^7$ &$2.6\times10^4$&$4.5\times10^5$&0.0038\\  %
\hline 
Advanced & 6.58 &0.022 & 0& 0& 0 & 2.56 \\ %
\hline\hline 
\end{tabular} 
\caption{The cross sections (in unit of fb) of the signal and SM background processes after each cut, 
and the corresponding significance  
${\cal S}=S/\sqrt{S+B}$. The integrated luminosity is  $\cal{L}=$1 fb$^{-1}$ and the 
running energy is $\sqrt{s}=4.18 \GeV$. 
We choose $m_\chi=0.1 \GeV$ and $\varepsilon=0.01$ 
for the millicharged particle. 
The advanced cut here is $E_\gamma/\text{GeV}=0.99 z_\gamma^2 + 0.99$.
}
\label{tab:cuts}
\end{center} 
\end{table} 

We optimize the advanced cuts for each BESIII running energy 
by choosing the $a$ and $b$ values that maximize the significance 
for the case in which $m_\chi=0.1$ GeV and $\varepsilon=0.01$.
Four SM backgrounds are considered in the optimization, 
including 
$e^+ e^- \to \nu\bar\nu\gamma$, 
$e^+ e^- \to e^+e^-\gamma$, 
$e^+ e^- \to \gamma\gamma\gamma$,  
and the reducible background in meson decays.
Table (\ref{tab:data}) shows the optimized $a$ and $b$ values for each 
running energy at BESIII. 
We have checked that under the optimized advanced cuts, 
the $e^+ e^- \to \mu^+ \mu^-  \gamma$ process 
does not contribute any background event.


\begin{table}[h] 
\begin{center} 
\begin{tabular}{|c|c|c|c|c|c|} 
\hline 
\hline
Year& $\sqrt s$ (GeV)& ${\cal L}$ (fb$^{-1}$) & $a$ & $b$  & $\varepsilon_{95}$\\
\hline 
2015 & 2.125 & 0.1      & 0.52 & 0.53 & 0.015  \\%
2012 & 3.097 & 0.32    & 0.68 & 1.12 & 0.015  \\%
2017 & 3.515 & 0.5     &  0.79 & 0.86 & 0.0095  \\%
2011 & 3.554 & 0.024  & 0.84 & 0.86 & 0.044  \\%
2012 & 3.686 & 0.51    & 0.95 & 1.21 & 0.013  \\%
2011 & 3.773 & 1.99    & 0.89 & 0.94 & 0.0051 \\%
2017 & 3.872 & 0.2     &  0.90 & 0.96 & 0.016   \\%
2011 & 4.009 & 0.5      & 0.92 & 0.98 & 0.011  \\%
2016 & 4.18   & 3.1      & 0.99 & 0.99 & 0.0060 \\%
2013 & 4.23   & 1.05    & 1.00 & 1.01 & 0.011  \\%
2013 & 4.26   & 0.83    & 1.01 & 1.02 & 0.013  \\%
2017 & 4.28   & 3.9     &  1.04 & 1.04 & 0.0063  \\%
2012 & 4.36   & 0.5      & 1.06 & 1.05 & 0.019  \\%
2014 & 4.42   & 1         & 1.02 & 1.08 & 0.014  \\%
2014 & 4.6     & 0.5      & 1.04 & 1.14 & 0.024  \\%
\hline 
11-17 & -    & 15.024     & - & - & $8.6 \times 10^{-4}$  \\%
\hline\hline 
\end{tabular} 
\caption{The center-of-mass energy and corresponding luminosities collected since 2011 
at the BESIII detector, and the corresponding optimized $a$ and $b$ parameters for the advanced cuts. 
$\varepsilon_{95}$ is the 95\% C.L.\ upper limit on millicharge $\varepsilon$ for the 
$m_\chi=0.1$ GeV case. Data before 2017 are given by \cite{database}, 
and information about 2017 data is provided by \cite{besdata2017}. 
The last row shows the limit combining all data 
between 2011 and 2017.}
\label{tab:data} 
\end{center} 
\end{table} 

{\it Methodology of combining data.}
A large amount of data have been accumulated by the BESIII 
detector at various running energies since 2011 
when the monophoton trigger was implemented 
\cite{trigger}. 
A summary of the BESIII data is presented in  
Table (\ref{tab:data}) 
where the data are arranged by the center-of-mass energy $\sqrt{s}$. 
To probe the millicharge, we  
carry out a likelihood analysis to combine all the data 
collected at various running energies as 
shown in Table (\ref{tab:data}). 
We first define a chi-square at each running energy 
\be
\chi_i^2 = {S_i \over \sqrt{S_i + B_i} },
\ee
where $S_i$ ($B_i$) is the number of signal (background) events 
at the running energy labeled by the index $i$. 
Here $B_i$ includes the SM irreducible background, the 
SM reducible backgrounds, 
and other possible background events caused by 
instruments. 
We further 
build a likelihood function ${\cal L}_i$ for each running energy as follows 
\be
{\cal L}_i = \exp{(- {\chi_i^2/2})}. 
\ee
The total likelihood function ${\cal L}$ for combing all the running energies can be 
built via 
\be
{\cal L} = \Pi_i w_i {\cal L}_i, 
\ee
where $w_i$ is the weight for each running energy. 
The test-statistic (TS) is related to the total likelihood via 
\be
\text{TS} = -2 \ln {\cal L}. 
\ee
The 95\% confidence level (C.L.)  exclusion limit on the millicharge $\varepsilon$ is obtained by 
demanding that the corresponding TS is larger by $2.71$ than 
that in SM. 
In our analysis, we set $w_i=1$ for all data points.


\begin{figure}[htbp]
\vspace{0.2cm}
\centering
\hspace{-0.6cm}
\includegraphics[scale=0.45]{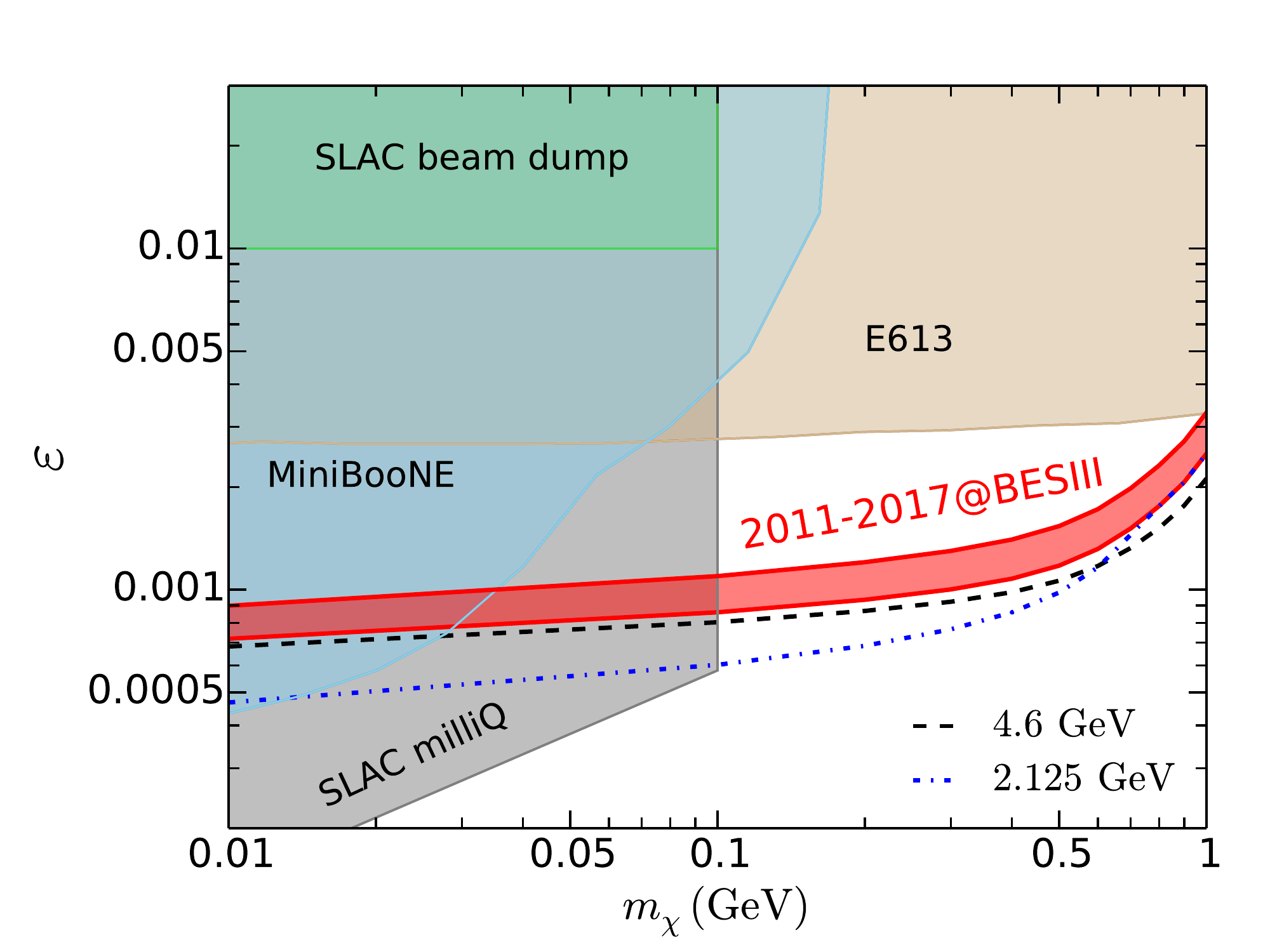}
\caption{
The expected 95\% C.L. exclusion limits on millicharged 
particles using the BESIII data collected during 2011-2017. 
The upper (lower) edge of the red band corresponds to the case where 
five (zero) more events appear in the BESIII data which 
are caused by instruments. 
The black-dashed (blue-dotdashed) line is the projected limit 
by assuming an additional 10 fb$^{-1}$ data 
in future BESIII runs with $\sqrt{s}=4.6\, (2.125)$ GeV.
Existing bounds are shown as shaded regions: 
bounds from the SLAC electron beam-dump experiment \cite{Davidson:1991si}, 
bounds from the SLAC MilliQ searches \cite{Prinz:1998ua}, 
bounds from E613 \cite{Soper:2014ska}, 
and bounds from MiniBooNE \cite{Magill:2018tbb}.
}
\label{fig:result}
\end{figure}

{\it Results.}
Fig.\ (\ref{fig:result}) shows the combined 95\% C.L. exclusion upper limits on millicharge  
$\varepsilon$ for various masses, by using the data presented in Table (\ref{tab:data}). 
Here we further consider other possible background events that are caused by 
instruments; 
a simple Monte-Carlo algorithm is used to assign these instrumental background events to 
various running energy points according to the {integrated} luminosity. 
As shown in Fig.\ (\ref{fig:result}), the current BESIII data can probe millicharge 
down to $\varepsilon \lesssim 0.86\, (2.5) \times 10^{-3}$ 
for the $m_\chi = 0.1\, (1)$ GeV case. 
 Fig.\ (\ref{fig:result}) further shows 
pre-existing experimental constraints 
in the $0.01~\text{GeV} \lesssim m_\chi \lesssim 1~\text{GeV}$ mass range, 
which include 
bounds from the SLAC electron beam-dump experiment \cite{Davidson:1991si}, 
bounds from the SLAC MilliQ searches \cite{Prinz:1998ua}, 
bounds from E613 \cite{Soper:2014ska}, 
and bounds from MiniBooNE \cite{Magill:2018tbb}.
Thus, BESIII data can provide new leading upper limits  
to the millicharged particle in the mass range, 
$0.1~\text{GeV} \lesssim m_\chi \lesssim 1~\text{GeV}$. 
The new limits significantly reduce the parameter space 
in which one can use 1\% millicharged DM to 
explain the 21 cm anomaly 
\cite{Munoz:2018pzp, Berlin:2018sjs, Barkana:2018qrx}.

Projections are made under the assumption that {additional 
10 fb$^{-1}$ data} are to be collected in the future BESIII runs. 
Two different projected limits are drawn on Fig.\ (\ref{fig:result}) 
for collecting 10 fb$^{-1}$ more data at 
$\sqrt{s}=2.125$ GeV 
and $\sqrt{s}=4.6$ GeV; 
realistic data takings with more running energies can be 
interpolated between these two projected limit lines.



{\it Summary.}
In this work, we have proposed a search for {millicharged} particles 
via the monophoton signature at the BESIII detector at BEPCII. 
We found that by using the current BESIII data, one 
can provide new leading constraints 
on the millicharged particle in the mass range, 
$0.1~\text{GeV} \lesssim m_\chi \lesssim 1~\text{GeV}$. 
We also systematically analyzed the irreducible and 
reducible SM backgrounds for the 
monophoton signature that are essential for dark 
matter searches at the BESIII detector  
which was lacking in the literature to our 
knowledge.

{\it Acknowledgements.}
We thank 
Qing-Hong Cao, 
Junmou Chen, 
Shenjian Chen, 
Min He,
Shan Jin,  
Dayong Wang,
Qi-Shu Yan, 
Peng-Fei Yin, 
Zhao-Huan Yu,  
Jianlei Zhang, 
Lei Zhang, 
Ruilin Zhu 
for helpful correspondence
or discussions. The work is supported in part  
by the National Natural Science Foundation of China under Grant Nos.\ 
11775109 and U1738134, 
by the National Recruitment Program for Young Professionals, 
by the Nanjing University Grant 14902303, 
by the China Postdoctoral Science Foundation 
under Grant No.\ 2017M611771.



\begin{thebibliography}{99}
\bibitem{cut}
Reducing the soft photon cut to $E_\gamma>0.01$ MeV 
does not alter the advanced cuts 
nor the monophoton events in 
the $e^+ e^- \to \gamma \gamma \gamma$ process 
after the advanced cuts.
\end{thebibliography}
\end{document}